\documentclass{aa}  
\usepackage{natbib}
\bibpunct{(}{)}{;}{a}{}{,}
\usepackage{graphicx}

\begin{document}

\authorrunning{S. Criscuoli and I. Ermolli}
\titlerunning{Restoration of full-disk CaII K time series}
   \title{Stray-light restoration of full-disk  CaII K solar observations:\\ a case study}
   \author{S. Criscuoli and I. Ermolli}

   \offprints{Serena Criscuoli \email{criscuoli@oaroma.inaf.it}}

   \institute{ INAF - Osservatorio Astronomico di Roma, Via Frascati 33, I-00040, Monte Porzio Catone, Italia
   }
   \date{}

   \abstract
{} 
{We investigate whether restoration techniques, such as those developed for application to current observations,
can be used to remove stray-light degradation effects 
on archive CaII K full-disk observations. 
We analyze to what extent these techniques can recover homogeneous time series of data.}
{We 
develop a restoration algorithm based on a method presented by Walton \& Preminger (1999). We apply this algorithm to data for 
both present-day and archive CaII K full-disk observations, 
which were acquired using the PSPT mounted at the Rome Observatory, or obtained 
by digitization of Mt Wilson photographic-archive spectroheliograms.}
{We show that the restoring algorithm improves both spatial resolution and photometric contrast of the analyzed solar observations. 
We find that the 
improvement 
in spatial resolution is similar for analyzed recent and archive  data. On the other hand, 
the improvement 
of photometric contrast is quite poor for the archive 
data, with respect to the one obtained for the present-day images. We show that the quality of restored archive data depends on the photographic 
calibration applied to the original observations. In particular, photometry can be recovered with a restoring algorithm if the 
photographic-calibration preserves the intensity 
information stored in the original data, principally outside the solar-disk
observations.}
{}

\keywords{Sun: activity - Sun: chromosphere - Sun: faculae, plages - Techniques: images processing}

\maketitle

\section{Introduction}
     \label{S-Introduction} 

Long time series of synoptic full-disk CaII K observations are applied in many fields of solar
research, in particular  all investigations concerning solar activity and variability over long-term timescales \citep{solanki2007}. 
Among existing series, some include archive observations that were recorded over many decades. These series, 
in addition to those using present-day observations, are composed of data affected differently by degradations due to 
instrumental optics and turbulent motions in the Earth's atmosphere. 
The removal of these degradations enables higher-quality analyses and measurements of data stored in the series.

The literature concerning the estimation and correction of instrumental and atmospheric degradations of full-disk solar images is extensive
\citep[e.g.][]{brahde1972,hansen1973,albregtsen1981,lawrence1985,barducci1990,martinez1992}.
 \citet {toner1993} and \citet {toner1997} proposed a technique based on the estimate of the zero crossing of the Hankel transform for 
observed images.
Starting from these works, \citet{walton1999} presented a technique (WP, hereafter) based on the estimate of the Point Spread Function 
(PSF, hereafter) using the analysis of the Hankel transform of the images.
This technique was demonstrated to work on both synthetic and real solar observations.  
The authors 
recovered image photometry to an accuracy of less than a few tenths of a percent for CaII K observations taken at the San Fernando Observatory, 
and an increase in image homogeneity for the data set they analyzed. \citet{mathew2007}  reported similar results when they applied 
a modified version of the WP technique to remove instrumental contamination from analyzed time-series data of space-based SoHO/MDI full-disk continuum observations.

In this study we analyze whether, and to what extent, a restoring algorithm based on the WP technique can be used to remove 
instrumental and atmospheric degradations in archive CaII K observations. It is known that archive observations are affected by degradations 
which are caused by stray-light effects, that are stronger than those characteristic of  
present-day
data \citep{ermolli2007b}.  
With this aim, we developed a restoring algorithm that we applied to two different types of data. The 
first type of data are  
present-day 
observations acquired using the PSPT telescope located at the Rome Observatory. The second type of data that we analyzed included observations  
derived using 
digitization of 
the Mt Wilson solar photographic archive. At present, these data are the only archive CaII K observations  freely available in digital format
after  pre-processing, which includes the photographic-calibration of the original observations. The data used in this study are 
presented in  Sect. 2. The brief description of the restoring algorithm that we applied, in addition to an account of its implementation, are provided
 in Sect. 3. 
We present the results obtained  by examining the 
analyzed images in Sect. 4. We then discuss our results in Sect. 5 and our conclusions in Sect. 6.

\section{Data}
\label{Data}
Our results are based on the analysis of the four data sets of CaII K full-disk observations listed below.

{\it PSPT data set (PSPT, hereafter)-} 
This data set is made up of current observations
acquired using the PSPT telescope at the Rome Observatory \citep{coulter1994, ermolli2003}. These data  were
 obtained by a 2048 $\times$ 2048 12 bit pixel CCD camera, 2$\times$ binned, with a final pixel scale of $\approx$ 2 arcsec, using an 
 interference filter centered
on the CaII K line (393.3 nm, FWHM 0.25 nm). We analyzed 170 images obtained on different observing days during the months of  July to  September from 2000 to 2005.
The analyzed data are composed of single-frame images acquired using short-exposure times ($ <$ 50 ms) and are  
available from the Rome-PSPT archive\footnote{http://www.mporzio.astro.it/solare}.

{\it Single-Day PSPT data set (SD-PSPT, hereafter)-}
We  analyzed  {\it PSPT} images acquired at different times during a single observing day, namely June 12, 2003. 
In particular, we analyzed a set of 23 images acquired from 6:01 to 7:08 UT
every three minutes, and two sets of 14 images each, acquired every 15 minutes, from 7:29 to 10:44UT, and from 13:15 to 16:36 UT,
 respectively. Because of seeing variability during the observing day and the small evolution in the analyzed solar features during the 
 same period, the analysis of these images tested the algorithm's capability to restore images to a common standard.

{\it Mt Wilson-CAL1 data set (MW, hereafter)-}
This data set includes 290 images obtained by the digitization of the archive CaII K full-disk  spectroheliograms, centered on the CaII K line (393.3 nm with 0.02 nm spectral resolution), taken at 
the Mt Wilson Observatory \citep{ulrich2004,lefebvre2005} during years 1961, 1967 and 1975. These data are 
photographically calibrated FITS files, which are available from the UCLA 
archive\footnote{http://www.astro.ucla.edu/$\sim$ ulrich/MW$_{}$SPADP/}. Details about the calibration procedure applied to the original observations
can be found on the UCLA web-page. 
The {\it MW} images are 800 $\times$800 pixels, 16-bit gray-scale files,  
and the pixel scale is $\approx$ 2.7 arcsec/pixel. 
The file headers of these images contain information provided by the UCLA project scientists 
about the  original plate and its digitization, in addition to measurements of the solar-disk position and shape, namely 
center position,
horizontal, and vertical radii. 

{\it Mt Wilson-CAL2 data set (test-MW, hereafter)-}
We analyzed  50  photographically un-calibrated Mt Wilson observations acquired during the years 1967 and 1975, which correspond to 
50  {\it MW}
images. We calibrated these images by applying a method presented in the literature 
\citep{burkanian}. 
In brief, we converted the pixel values of the un-calibrated images  
into relative intensity values according to the 
formula: $I=(V-B)/(T-B)$, where $I$ is the intensity (in arbitrary units), $V$ is the average of pixel Data Number ($DN$, hereafter) values 
for the un-exposed part of the plate, $B$ is the average of $DN$ values for the  
darkest pixels, and $T$ is the $DN$ value for a given pixel. 
The criteria applied for the identification of both darkest and un-exposed pixels are the following. The first ones are all the  image pixels with
intensity equal to or lower than the minimum DN value of solar-disk pixels, while the pixels of the un-exposed part of the plate are the ones outside the solar-disk
with intensity higher than the maximum $DN$ value of solar-disk pixels.  
 The data obtained by this formula  are 800 $\times$800 pixels, 16-bit gray-scale images, and have pixel scale  $\approx$ 2.7 arcsec/pixel, as for the
 {\it MW} data set 
analyzed. 

\begin{figure}
   \centering{
 \includegraphics[width=7.7cm]{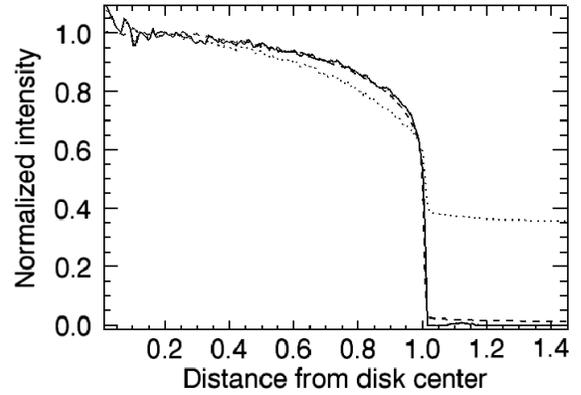}} 
       \caption{Profiles of the median intensity values computed inside constant area annuli centered on the solar-disk center in 
	    {\it MW} (solid line),  {\it test-MW} (dotted line), and {\it PSPT} (dashed line) images. Distance from the disk center is given in 
		disk radius units. Intensities are normalized to the value extrapolated at disk center from 0.3 solar radii. These profiles are computed from images storing observations
		 taken on December 29, 1975 for {\it MW} and {\it test-MW} data sets, and on August 31, 2005 for the {\it PSPT} sample.}
         \label{fig_profiles_calib}
   \end{figure}

Figure \ref{fig_profiles_calib} shows the profiles of the median intensity values computed on constant area annuli that are centered in the solar-disk center
in {\it MW}, {\it test-MW}, and {\it PSPT} images. We note the difference in the median intensity values for the solar limb and outside the solar-disk 
between {\it test-MW} and {\it PSPT} and {\it MW}  images. This is caused not only by the different level of stray-light, but also by the photographic-calibration 
procedure applied to the original Mt Wilson observations.
%

\section{The restoring algorithm}

The restoring algorithm that we developed is based on a technique presented by \citet{walton1999}. In brief, the intensity Center-to-Limb 
Variation of the quiet Sun (CLVq hereafter) and the PSF are modelled using analytical functions (namely Legendre polynomials for the CLVq, and a 
sum of three Gaussians and a Lorentzian for the PSF), which depend on free parameters. 
These parameters were estimated for each image by fitting the measured CLVq to the convolution between the 
analytical model functions.
 From the retrieved PSF an optimal Wiener filter \citep{brault} was computed, with which images were  finally restored. 
A complete description of the technique is found in the original paper by \citet{walton1999}. Here we discuss some important 
aspects of its implementation.

The WP technique is based on the assumption that both atmospheric and instrumental  distortions 
are uniform and radially symmetric over the focal plane. 
For what concerns the first kind of aberrations, this assumption is valid for all data sets because the integration time was longer than the typical atmospheric correlation time of 10-20 ms. 
\citet{walton1999}
 provide a deeper discussion of these assumptions for when atmospheric degradation is not space-invariant.
Concerning instrumental degradations, we evaluated
that the eccentricity of the solar-disk on {\it PSPT} and {\it SD-PSPT} images is sub-pixel. Radial symmetry is 
therefore a 
reasonable approximation for such data. In both {\it MW} and {\it test-MW} images, the
eccentricity is approximately three times larger than for PSPT images. The radial symmetry 
is recovered by resizing the images by an amount proportional to the difference between 
the values of the `horizontal' and `vertical' radii provided in the image header.
Compared to present-day observations, {\it MW} and {\it test-MW} images were more strongly affected by asymmetric artifacts, which were 
produced during observations and storage, 
for example large-scale inhomogeneity affecting solar-disk observations and small-scale emulsion scratches.
The effects of such inhomogeneity on the quality of the restoring achieved for the analyzed data, is discussed in Sect. 4.

In agreement with results presented by \citet{walton1999} (see their Fig.1), we found that the convolution of the chosen model functions 
reproduced the observed solar profile in both {\it PSPT} and {\it MW} images. The typical value of the chi-square statistic was approximately 
$10^{-4}$ for both data sets. The functions that describe the CLVq and the PSF were not however orthogonal; moreover the best-fit merit 
function had several minima and the values of the coefficients were dependent on the initial guess values 
(see next section). To investigate these issues we applied the developed algorithm to synthetic images of the Sun convolved by various PSFs. We found 
that the fitting produces large variations in the coefficient values that describe the 
CLVq, in some cases even when the coefficients of the CLVq were initialized to the values used in the simulation. 
This was because the terms of the Hessian which correspond to the coefficients of the PSF were smaller than the terms corresponding to the coefficients of the CLVq. 
We found in addition that the effect increased as the aberration increased.
This may cause a poor estimation of the aberrations and a poor restoration of the images. 
In order to reduce this effect, we carefully estimated the first guesses of coefficient values.
The coefficients that describe the CLVq for {\it PSPT} images
were initialized using values obtained 
by fitting intensity profiles for good quality images acquired at minimum activity level to the model CLVq function. This preliminary fit 
was repeated for each {\it MW} image, 
because the large spread in intensity profiles did not allow a single set of coefficients to be 
assumed for all of the data analyzed.
The set of coefficients that describes the PSF was initialized using  values obtained by performing the fit on 
profiles evaluated for images that were randomly selected from all data sets.
The radius of the solar-disk image was also a free parameter of the fit. To estimate a first guess, we applied a method based on the application of an 
edge-enhancement operator 
to solar images: The edge-enhancement operator was applied using an intensity threshold, the solar-activity features were removed, and the mean of the maximum diameters in 
the horizontal and vertical directions were computed.
We note that the use of values 
derived using such a method, enables a best-fit function to be found more efficiently, compared to the use of values of radii provided in 
{\it MW} and {\it test-MW} image headers, which usually over-estimate our radius measurements.

The algorithm we developed differs from the technique described by \citet{walton1999} in the following aspects:

\begin{itemize}
    \item \textbf{The model function for the CLVq}: WP employed a 3th-order Legendre polynomial expansion. Following \citet{pierce} we employed a 
	5th-order polynomial expansion in $\mu$. 
\item\textbf{Estimation of CLVq on images}: WP estimated the CLVq by computing the median intensity value on concentric constant-area annuli 
of the solar-disk, using the standard deviation as an estimate of the error. In our algorithm, we define a quiet Sun intensity value to be the $5\%$ of the maximum value of the cumulative intensity histogram in each annulus \citep{brandt}.
The error was evaluated by taking into account the intensity values that  
corresponded to $3\%$ and $7\%$ of the maximum value of the cumulative intensity histogram within each annulus; we 
defined the error to be the largest variation between these values and the one estimated for quiet Sun regions. We found that 
this method discarded active regions more efficiently, in particular for images acquired at maximum activity level. The CLVq  calculated using 
this method extends to approximately 1.4 and 1.5 solar radii in {\it PSPT} and {\it MW} images, respectively.
 \item \textbf{Minimization}:
WP modelled analytically the Hankel transforms of the PSF and the CLVq, computed numerically 
 the inverse Hankel transform of their product, and fitted the experimental CLVq using the resulting vector. In our algorithm,  we  computed the 
 model bi-dimensional images representative of the PSF and the CLVq over the solar-disk, then computed the product of their Fourier Transforms (FTs), 
 anti-transformed, and finally fit the data using the profile of the resulting model image. This meant that our algorithm 
became  slower than that of the original WP method, because it required the computation of 3 FTs of bi-dimensional images at each iteration of the
 minimization, 
in contrast to a single numerical computation of the inverse Hankel transform of a vector. Nevertheless, we decided to compute FTs, because
 previous works 
 showed that the Hankel transform numerical computation is `inherently less accurate' \citep{walton1999} than that of the FT. Moreover, 
 the computation time for application of our algorithm to the entire sample of data considered in this study appears reasonable 
 (see details below). We note that in a similar way to the WP method, no FT of real data is computed for minimization. 
\item \textbf{Second fit}: to improve the estimation of the PSF, a second fit was performed only on and outside the limb keeping fixed the values estimated for the CLVq. 
In particular, for {\it PSPT} images the second fit was performed for the range 0.95-1.4 solar-disk radii. 
Because of the calibration procedure, the lowest intensity values for {\it MW} images out of the disk were set to zero, which allowed the fit to be performed for the range 0.95-1.07 solar disk radii.
\end{itemize}

The restoring algorithm we developed was implemented using the IDL language. The minimization, the bulk of the algorithm, was performed using the 
Levenberg-Marquardt fit subroutines of the MARQWARDT library \footnote{These software are available at Markwardt IDL Library \\http://cow.physics.wisc.edu/$\sim$craigm/idl/idl.html}.
The restoration of each image, performed by a PC with 1.73GHz processor and 2G Ram, took from 5 to 30 minutes for all the data sets. 

\section{Results}
Images from the {\it SD-PSPT}, {\it PSPT}, and {\it MW} data sets were restored using the implemented algorithm. The coefficients that  
describe the CLVq and the PSF were initialized with the same values for all images in a given data set. An example of original and restored images is 
given in the appendix.

We visually inspected the restored images and found that some were affected by artifacts, that is medium and high spatial frequency 
features which were not present in the original images, were apparent. 
These images were discarded and in subsequent analyses only 100$\%$, 75$\%$, and 55$\%$ frames from the {\it SD-PSPT}, {\it PSPT}, and {\it MW} data sets, 
respectively, were retained.
The presence of artifacts in restored images was due to the failure of the minimization algorithm to find a good set of parameters 
to describe the PSF, which in turn, as explained above, led to the poor initialization of the parameter values. 
Not surprisingly the largest number of retained frames was from the  {\it SD-PSPT}, which is the most homogeneous of the analyzed data sets 
in terms of image content and the observing conditions. For the {\it PSPT} data set, it is possible to derive  
restored images without artifacts by initializing the free parameters of the PSF using different values. For {\it MW} images,
this was not always possible because the convergence of the solution without artifacts was  hampered by the presence of defects in the original images, such 
as straight lines, scratches, double images and large-scale gradients.

A visual inspection of the  original and restored images showed that the application of the
restoring algorithm yielded an  
improvement in both spatial resolution and contrast of solar features identified in the analyzed images. In particular, restored images showed fine details in both
quiet and active regions. These details resembled ones observed in PSPT images taken at the Mauna Loa Solar Observatory in the best
seeing conditions. 
To evaluate the image-quality improvements
and test the capability of the technique to recover homogeneity in the analyzed data sets, 
we performed several measurements. The results obtained are presented below.
We note that, due to the different instrumentation, spectral range, observations storage and data reduction, a comparison of absolute values 
of measurements derived using images from the present day and archive data sets considered in this work is impossibile. For this reason we now 
compare the relative variations in the quantities measured using original and restored images for the three data sets.

\subsection{Image spatial scale}
\label{spasca}

   \begin{figure}
   \centering{
 \includegraphics[width=7.3cm]{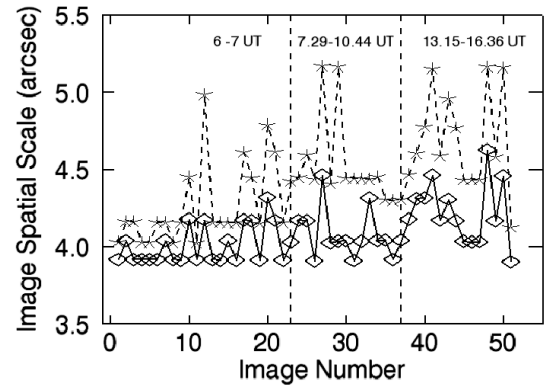}} 
       \caption{Evolution during the day of image spatial scale, estimated as described in the text, for original (dashed line, asterisks) 
	   and restored (solid line, diamonds) {\it SD-PSPT} images.}
         \label{fig_plot_resolution_oneday}
   \end{figure}
%

   \begin{figure}
   \centering{
 \includegraphics[width=7.5 cm]{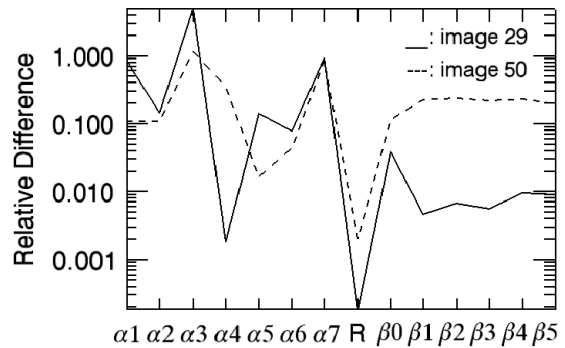}}
      \caption{Relative differences, in logarithmic scale, of the values of the coefficients estimated on images 29 and 50 of {\it SD-PSPT} sample
	  with respect the coefficients estimated for image 1  of the same sample. The set \{$\alpha$\} refers to the coefficients which describe the 
	  PSF, the set \{$\beta$\} refers to the coefficients which describe the CLVq. $R$ is the value of the solar-disk radius.}
         \label{fig_plot_resolution_oneday_coe}
   \end{figure}
%
We first analyzed the smallest spatial scale measured in each image. This was achieved by analyzing the power spectra of small sub-arrays
(64$\times$64 pixels) 
extracted at the disk center of each image. In particular, we assumed that the characteristic spatial-scale of each image, was the spatial frequency at which 
the integral of the power spectrum, over frequencies, is 98\% of its total value.
Figure \ref{fig_plot_resolution_oneday} shows the temporal variation of the measured spatial scale for original and restored images 
of the {\it SD-PSPT} sample. 
Restored images have better resolution, which in some cases approaches the Nyquist frequency (3.8 arcsec). Moreover, less scatter between 
the measured 
 values was 
found in the restored data set with respect to the original one.
In particular, the increase with time in the characteristic spatial scale of the original images, which reflects the deterioration of observing 
conditions during the observational day, was less evident for restored images. This shows that the 
algorithm works well for images of different quality.
Our most superior results, however,  were derived for images that had the highest spatial resolution.
The mean spatial scale measured for the 
first set of 23 images, which were acquired in the early morning, was
$4.3 \pm 0.3$ and $4.0\pm0.1$ arcsec for the original and restored data, respectively. On the other hand, the corresponding mean values for the 
14 images of the last data set, which were acquired during the central hours of the day, were are 
$4.7\pm0.3$ and $4.2\pm0.2$ arcsec, respectively. Exceptions did however exist, such as images  29 and 50, for which we measured the 
same spatial scale 
on  original images, but different values for the restored data. 
 We note that the mean resolution for the entire data set varies from $4.4\pm0.3$ to $4.1\pm0.2$ arcsec, which 
corresponds to an increase of 7$\%$ in the resolution, and a decrease in the data dispersion of approximately 33$\%$.

An inspection of the coefficients estimated by the minimization algorithm revealed that, for most images acquired in worse 
seeing conditions, the fitting produced large variations in the values of the coefficients that describe the CLVq.   
Figure \ref{fig_plot_resolution_oneday_coe} shows  for example the relative variations of the coefficient values estimated for images 
29 and 50, with respect to the values estimated for image 1 of the same data set. This latter image is one of the highest quality images obtained during  
that observing  day. The plot shows that for both images 
29 and 50 the largest variations are measured for the coefficients that describe the PSF, as expected. For image 50, large variations are 
measured also for the coefficients that describe the CLVq. 
This is due to the (already-mentioned) non-orthogonality of the PSF and CLVq functions and to the dependence of the fit on the initial conditions,
 which produces the remaining scatter of results measured in the restored data. For each image analyzed, we note that variations in the coefficient values that describe the PSF, are larger 
than those measured for coefficients that describe the CLVq; this indicates that the fit provides a reasonable approximation to the functions.
We note also that the variations in the values of each parameter are not directly related to the goodness of the restoration, which, in turn, depends on the goodness of the estimation of the `shapes' of the PSF and the CLVq functions. 

\begin{figure}
   \centering{
   \includegraphics[width=7.3 cm]{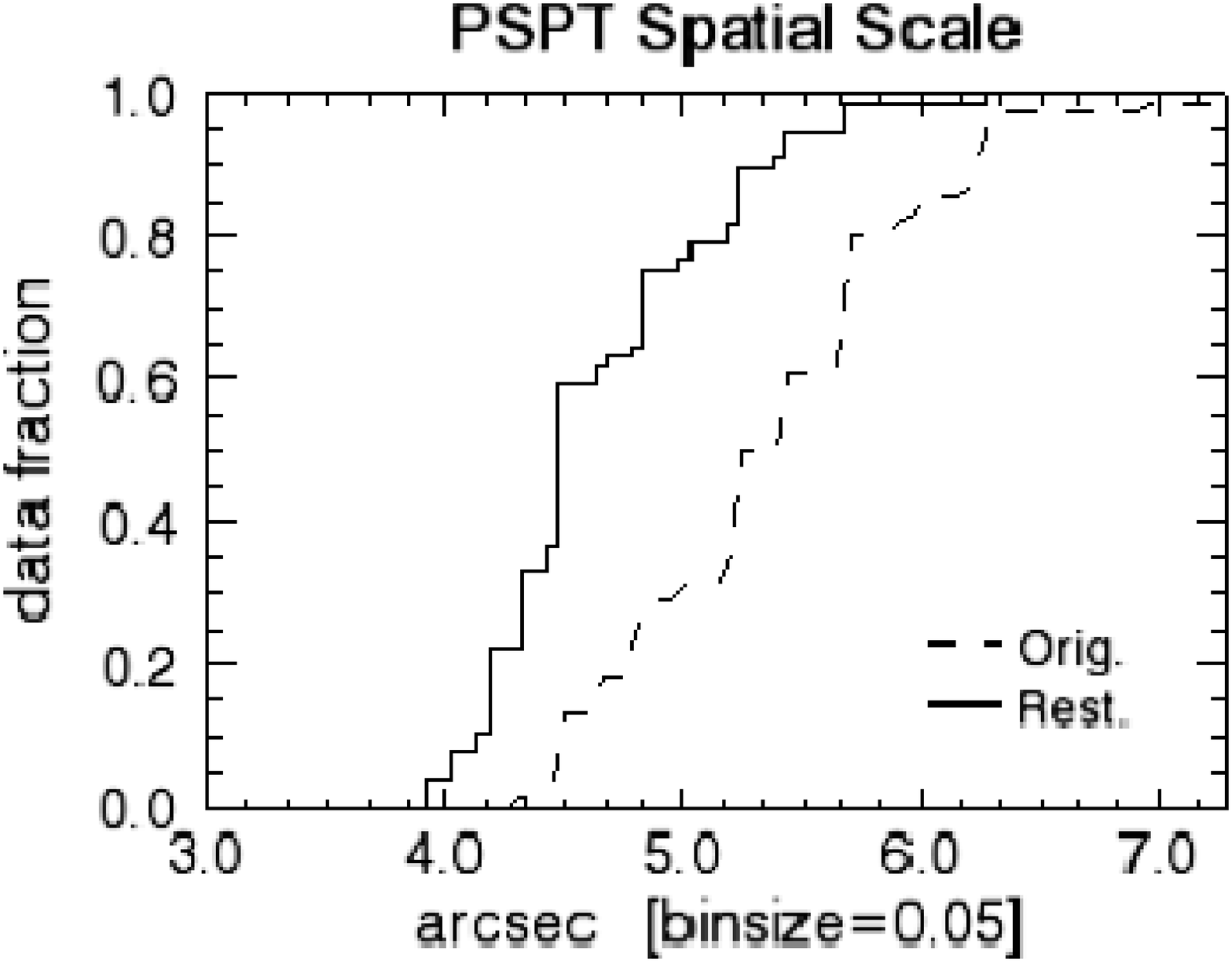}
   \includegraphics[width=7.3 cm]{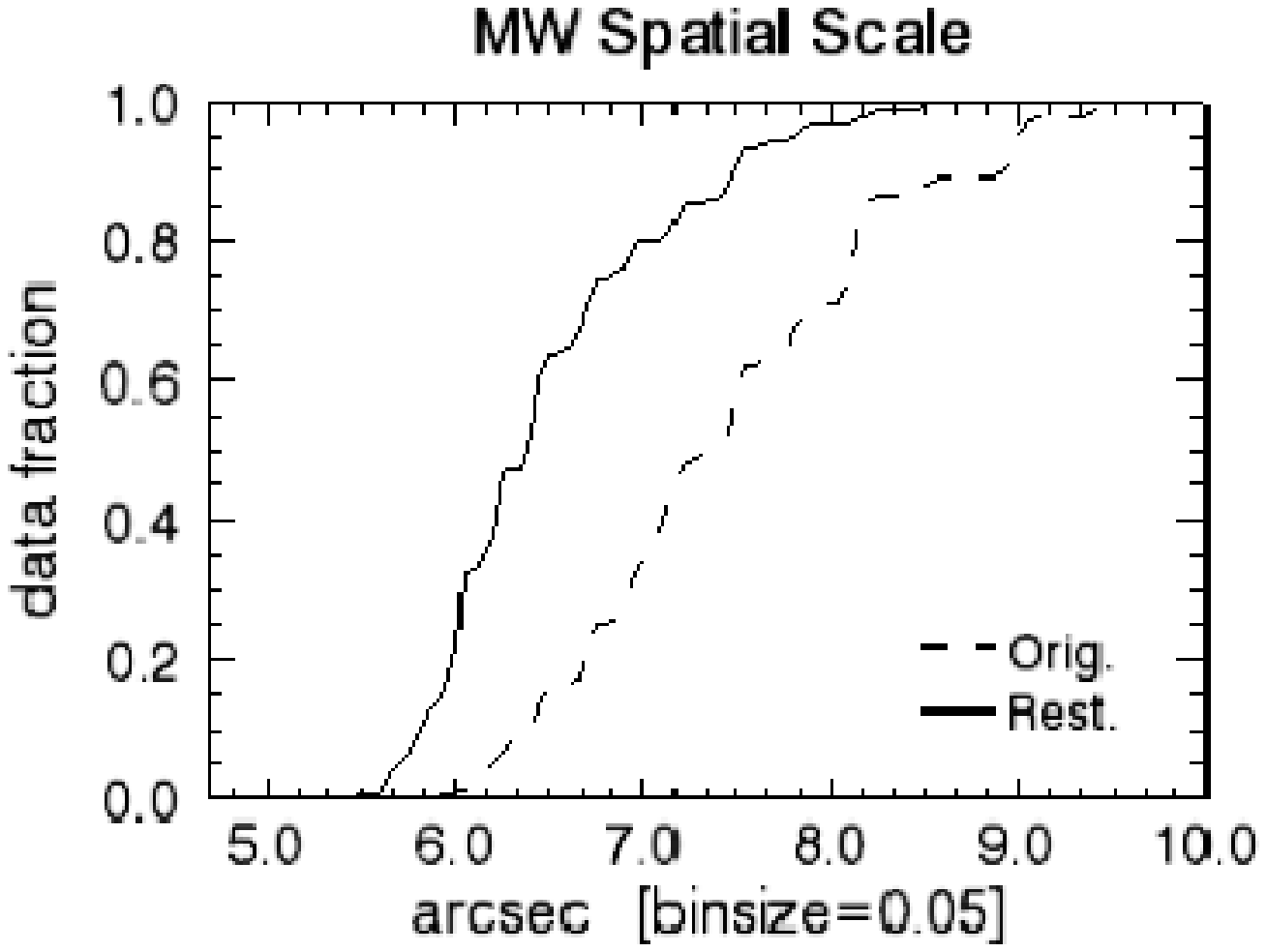}}
      \caption{Cumulative histograms of image spatial scale measured on restored (solid line) and 
	  original (dashed line) images.}
         \label{fig_plot_res_both}
   \end{figure}

Figure \ref{fig_plot_res_both} shows the cumulative histograms of the measured image resolution for original and restored images of the 
 {\it PSPT} and {\it MW} data sets that we analyzed.
These  plots clearly show that restoration improves the image resolution and homogeneity of both  analyzed data sets. In particular, the mean value of the measured spatial scale decreases from 
$5.4\pm0.7$ to $4.7\pm0.6$ arcsec and from $7.5\pm0.9$ to $6.5\pm0.7$ arcsec, for original and restored data of the PSPT and {\it MW} data sets, 
respectively; we note that this corresponds to an increase in resolution of approximately $13\%$ on both data sets.
The remaining scatter of the results measured after restoration, must be partially attributed to the non-orthogonality of the 
fitting functions, in a similar way to the {\it SD-PSPT} data set. 
The seasonal variation of the image scale and the variation of the image contents, in terms of solar-activity features occurring over the 
solar-disk, also contribute to the dispersion of the results. 
 
%
   \begin{figure}
   \centering{
 \includegraphics[width=7.3 cm]{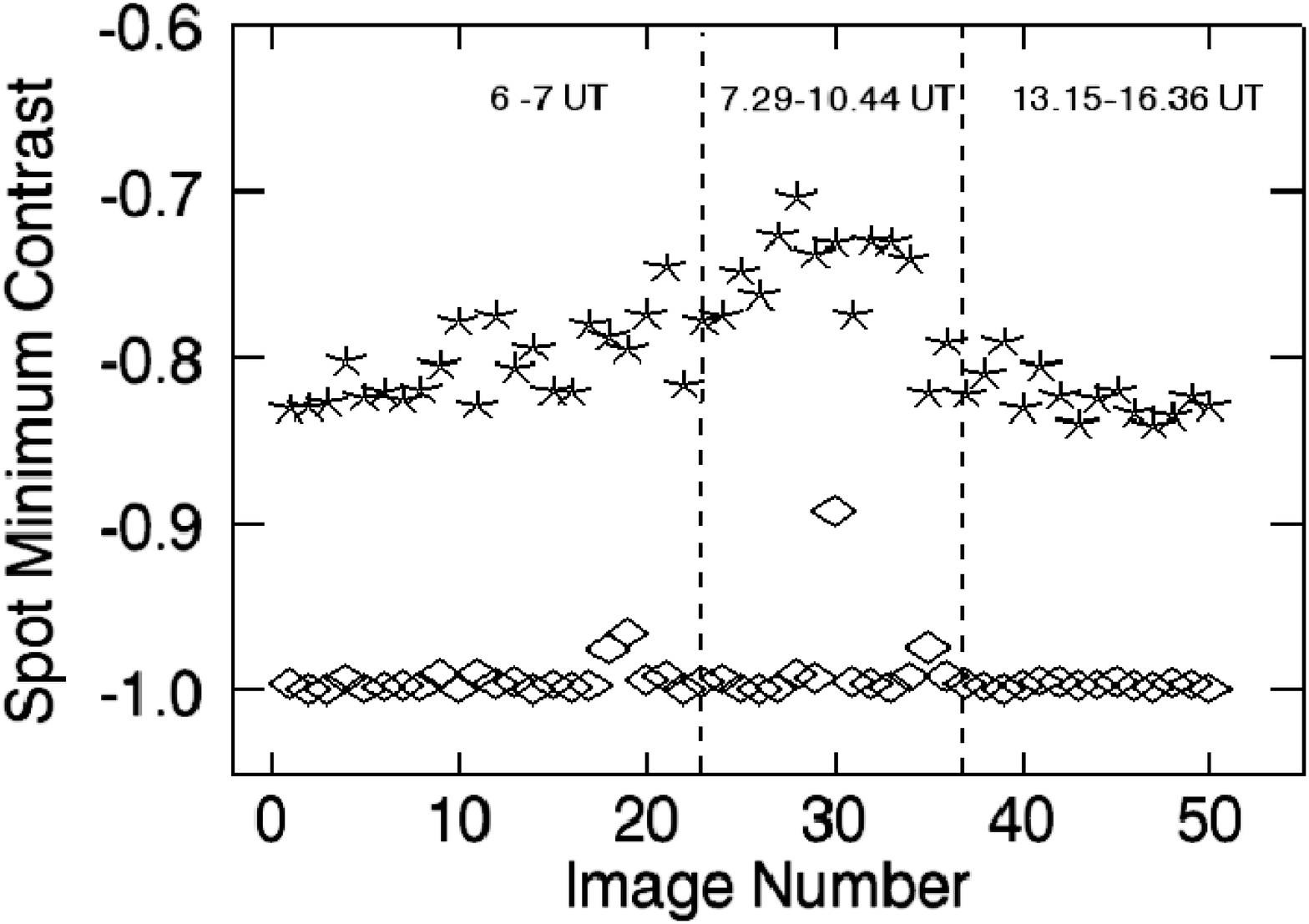}
}
      \caption{Evolution during the day of the minimum contrast of the largest sunspot (AR10380) measured on original (asterisks) and restored 
	  (diamonds) {\it SD-PSPT} images. The restoration decreases the contrast of approximately 30\% and reduces variations due to seeing variability.}
         \label{fig_plot_minimt_oneday}
   \end{figure}
%
 
\subsection{Contrast of solar features}
To study  the improvement in the photometric properties of solar features by restoration, we analyzed the contrast of sunspots and faculae 
identified for original and restored data. In the following, the contrast of each pixel is defined by the ratio of its own intensity and the 
intensity of quiet Sun region close by, the obtained quotient was then subtracted by one. The quiet Sun is established to be the median intensity 
value computed in annuli centered on the disk center.

{\it Sunspot regions -}
Sunspot regions were identified in the analyzed images by applying an  intensity-threshold criterion. We note that the application of this straightforward criterion 
for feature identification is functional to our  study, which takes into account only minimum contrast values measured for large-size 
identified features; we provide details of our procedure below.  
Figure \ref{fig_plot_minimt_oneday} shows the temporal variation of the minimum 
contrast measured for the largest 
sunspot identified on original and restored {\it SD-PSPT} images. The umbral area measured for the sunspot considered here, which belongs to 
AR10380 (shown in the online material), is approximately 70 micro-hemispheres.  
The plot shows a large dispersion in measured minimum contrast values for the original images. The maximum  dispersion was measured for
 images acquired during the hottest hours of the day (9:29 a.m. - 12:44 a.m. in local time units). 
The minimum contrast measured for restored images, however, has decreased by approximately $30\%$, and shows both less scatter and almost no trend during the whole observing 
day. 
   \begin{figure}
   \centering{
\includegraphics[width=7. cm]{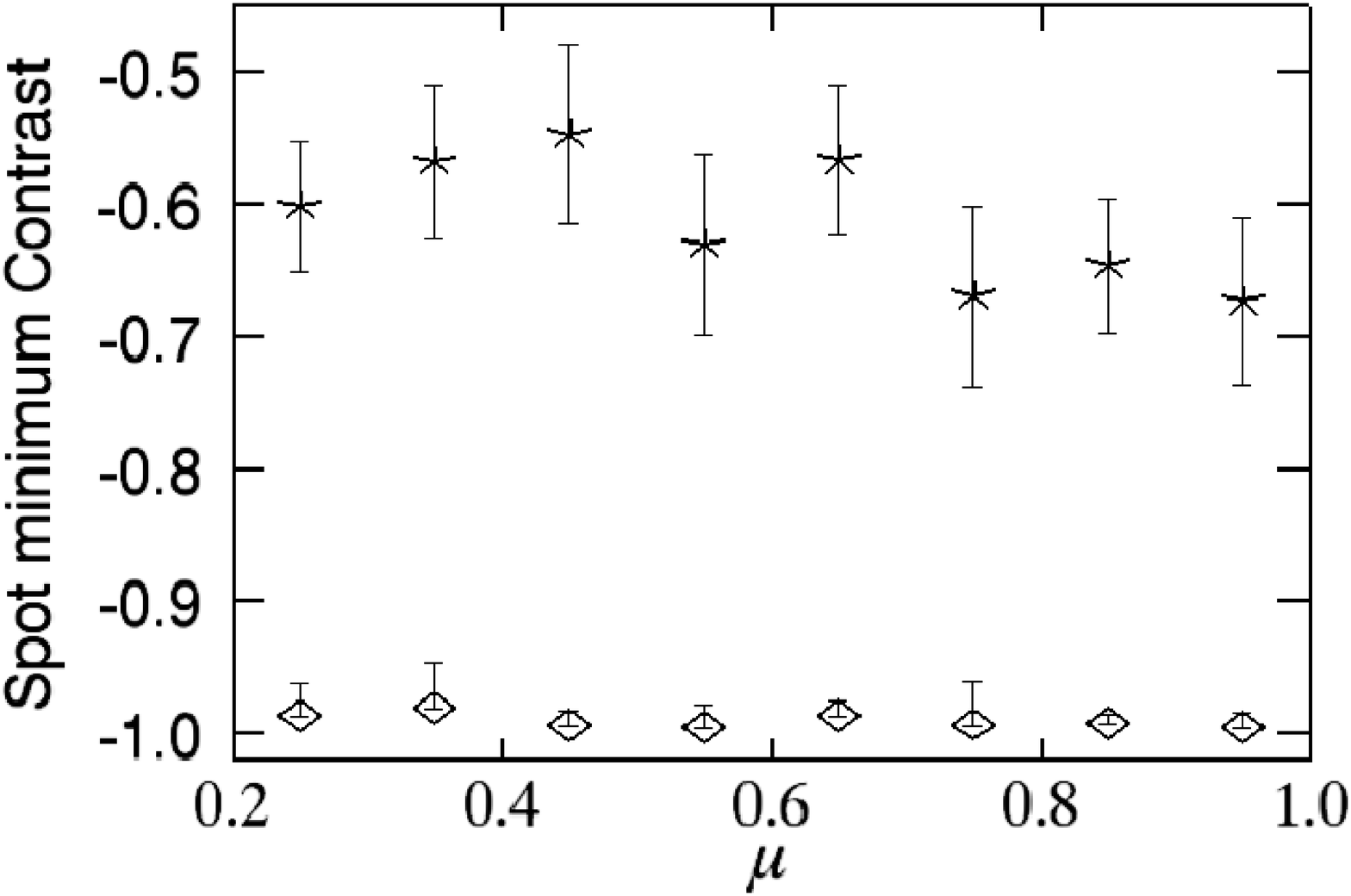}
\includegraphics[width=7.2 cm]{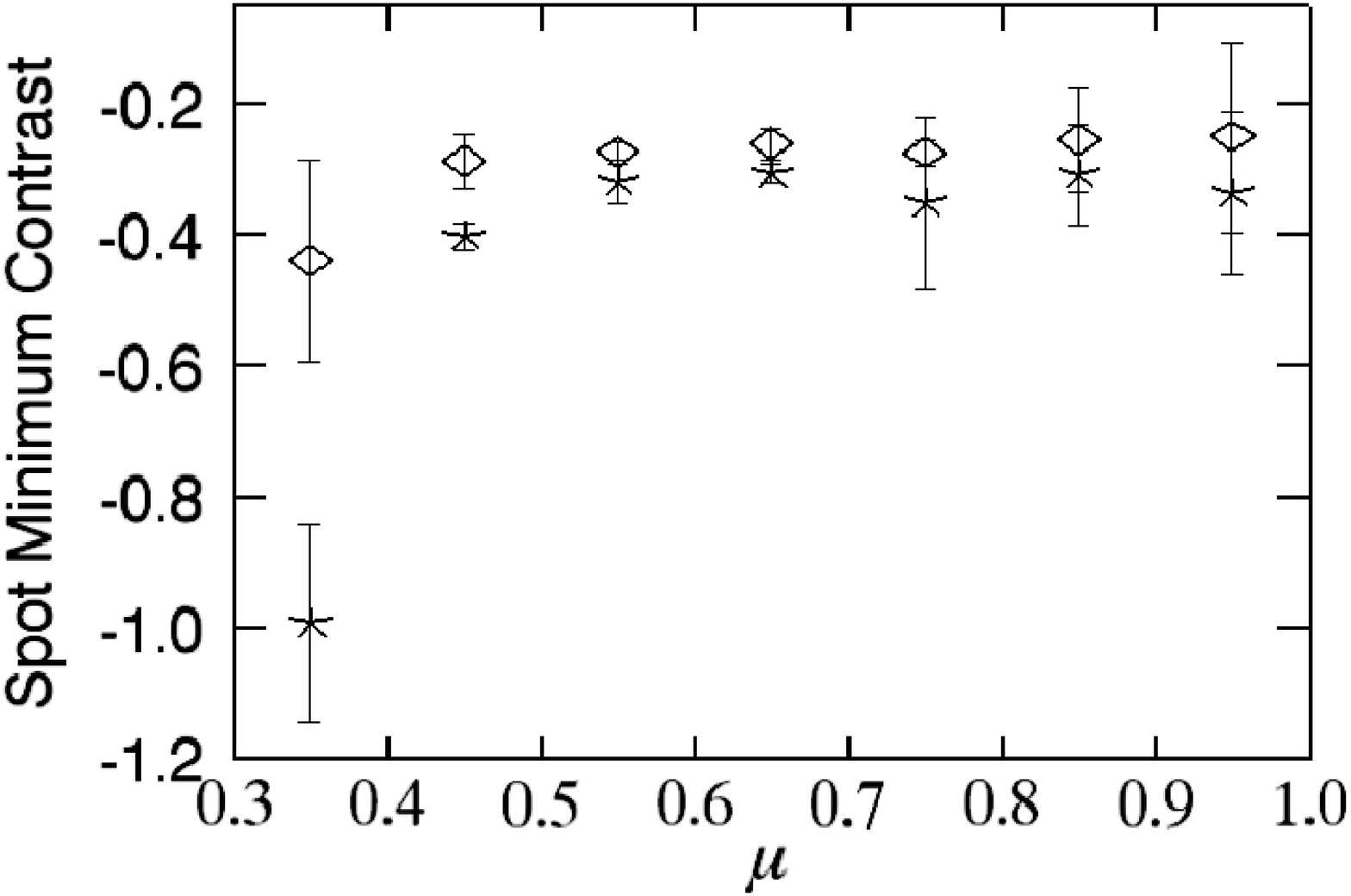}
}
      \caption{Center-to-limb variation of the minimum contrast of sunspots measured on original (asterisks) and restored (diamonds) images from {\it PSPT}  (top) and {\it MW} (bottom) data sets.}
         \label{fig_spotminint_both}
   \end{figure}
%

We analyzed the minimum contrast of sunspot regions identified in sub-samples of 76 {\it PSPT} and 61 {\it MW} images. To reduce the
potential dispersion in results associated with the dependence of measured contrast values on  sunspot size
 \citep{mathew2007,walton1999}, we restricted our analysis to identified features whose umbral area was larger than 30 micro-hemispheres. 
 We selected 233 and 57 on {\it PSPT} and {\it MW} images, respectively. We note that, due to the variation in the image contrast, 
 the number of selected features is larger after restoration (41$\%$ and 29$\%$  on {\it PSPT} and {\it MW} images, respectively), but  for these 
 analyses we retained only the features selected in both restored and original images.

In Fig. \ref{fig_spotminint_both} we plot the minimum contrast measured in the identified features against their position in the disk 
(in cosine of the heliocentric angle units). We show the value of the peak and the full-width-half-maximum (error bars) of the distribution of contrast values  inside 0.1 wide position
bins. We note that in the{\it PSPT} restored images the measured contrast of sunspots has decreased by  approximately 60$\%$ 
and the dependence on the position in the solar disk has reduced significantly with respect to the contrast measured in the original data.
Moreover, 
the uncertainties described by these error bars are smaller with respect to those derived for the original data, 
which provides additional confirmation that restoration improves the uniformity of the data sets. In contrast, restoration does not reduce the 
dispersion of the measurements 
 and the dependence on position within the solar-disk of the contrast measured in {\it MW} images. 
However, 
a small decrease in contrast values ($6\%$) is measured.  We note that the high (absolute) values of contrast measured in the restored {\it PSPT} 
images ($\approx 100\%$) are due primarly to the fact that
the sunspots analyzed here were observed in the CaII K spectral range. However, to estimate whether the obtained values stem from 
over-restoration of images, we applied the algorithm to a data set that was acquired simultaneously with the {\it SD-PSPT} data images 
for the red continuum spectral range (607.2 nm, FWHM=0.5nm).  
We measured sunspot-contrast values for restored images to agree with those presented by 
 \citet{mathew2007}. By analyzing simulated images of the Sun, which we describe in more detail below, we estimated that 
 contrast of sunspots of the size discussed is likely to be overestimated for low levels of stray-light.
 These findings agree with results presented by \citet{walton1999}. 

Finally, we investigated the contrast of sunspot features identified in {\it PSPT} images with 
umbral area values smaller than 30 micro-hemispheres. 
We found that the average value of the measured minimum contrast decreases with  restoring from -0.4 to -0.8. Moreover, the 
error bars associated with measurements performed on restored data were approximately twice as large as those derived for the original data, and a small trend 
with position on the solar-disk
was not removed by restoration. We  comment on these results in Sect. 5. 
   \begin{figure}
   \centering{
\includegraphics[width=7.2 cm]{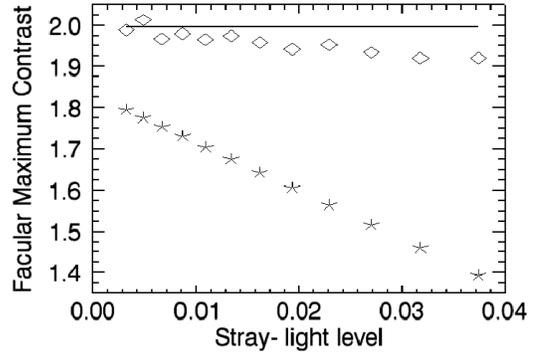}
}
\caption{Facular maximum contrast measured on simulated images versus the level of stray-light (see text). Continuous line: original simulated image. Asterisks: aberrated images. Diamonds: restored images.}
         \label{fig_simulaFAC}
   \end{figure}
%

   \begin{figure}
   \centering{
\includegraphics[width=7. cm]{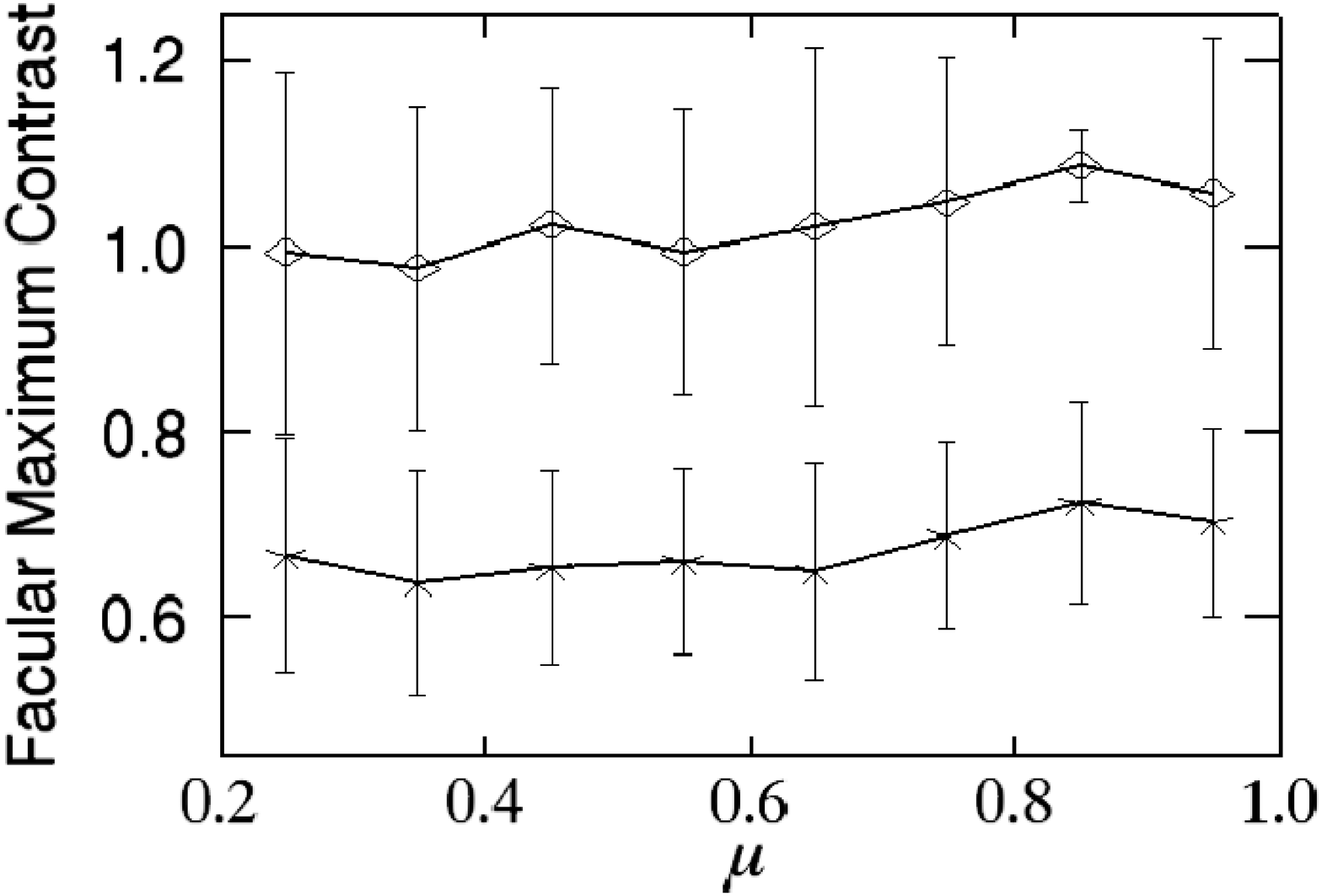}
\includegraphics[width=7. cm]{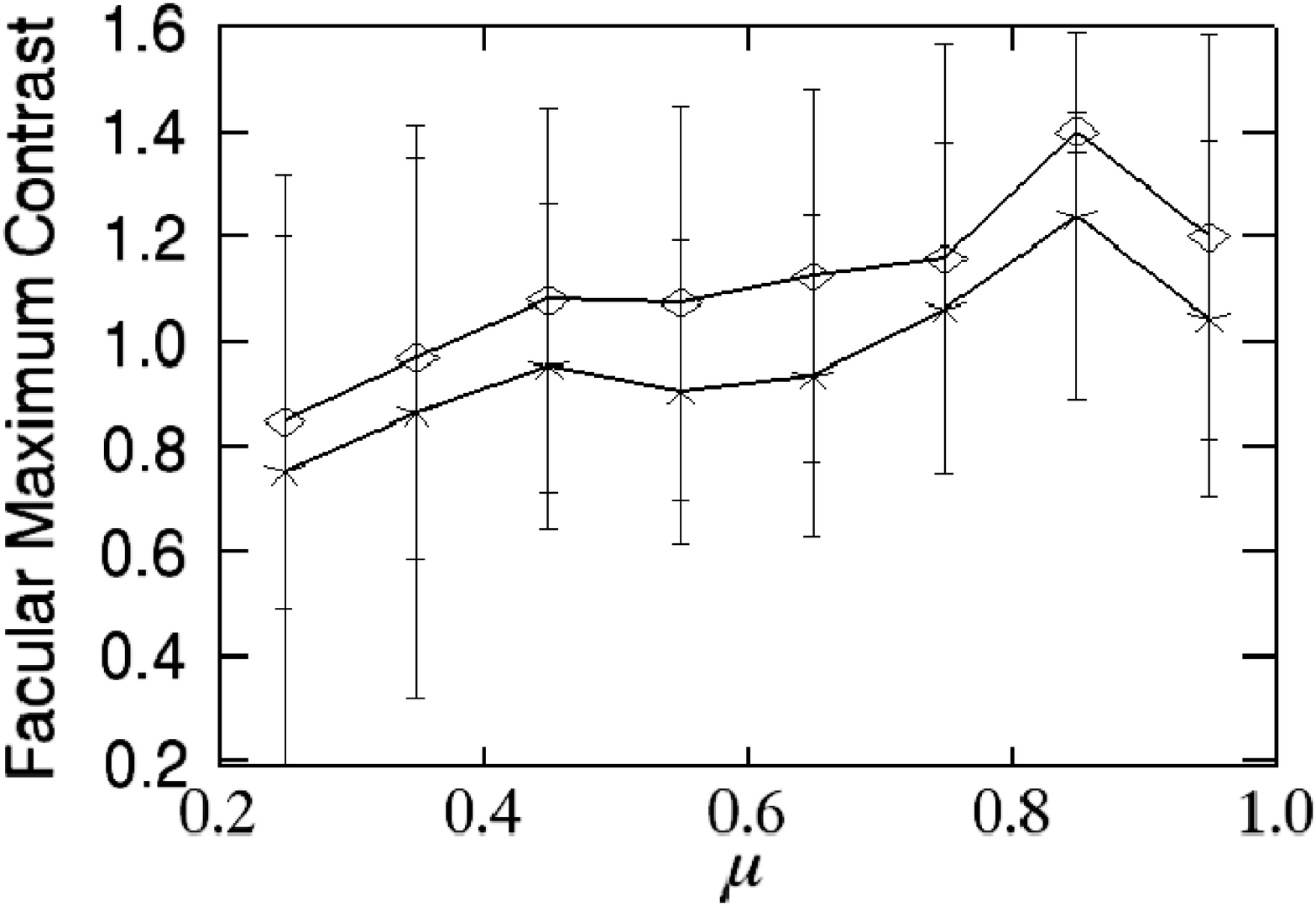}
}
\caption{Center-to-limb variation of maximum contrast of facular regions measured on original (asterisks) and restored (diamonds) images from {\it PSPT}  (top) and {\it MW} (bottom) data sets.}
         \label{fig_maxcontr_both}
   \end{figure}
%

{\it Facular regions -}
Facular regions were identified in {\it SD-PSPT} images and in {\it PSPT} and {\it MW} images used to study the contrast in sunspots. The 
identification method applied was Ktr, which was described by  \citet{ermolli2007}.
We found that the number of features identified in restored images was remarkably larger than those found in the original data, the increase being 
approximately 500$\%$, 300$\%$, and 35$\%$ for the {\it SD-PSPT}, {\it PSPT}, 
and {\it MW} data sets, respectively. 
The maximum contrast measured inside facular regions  also increased for the three data sets
of 33$\%$, 18$\%$, and 7$\%$, respectively.
We note that the maximum contrast measured inside facular regions  depends on the spectral characteristics 
of the analyzed images. In particular, the wider is the spectral range of the analyzed observations the lower is the contrast.

Since the AR1038 extended region observed in the {\it SD-PSPT}  sample developed several flares during the observing day, we could not study the temporal variations 
of  facular regions identified in restored and original images of that data set, as we did for 
sunspots. However, we studied the effects of the restoring algorithm on the  contrast of bright solar features 
by analyzing simulated images of the Sun.
We extracted a sub-array of 200$\times$140 pixels centered on the active region AR1038
from an image of the {\it SD-PSPT} acquired before the start of flaring activity. This sub-array was  superimposed on a simulated image of the Sun 
computed using  a polynomial 
expansion of $\mu$. The values of the coefficients were similar to those assumed to initialize the fit in {\it PSPT} images (see Sect. 3). The
derived image was convolved by twelve different PSFs corresponding to increasing 
levels of stray-light, which was defined to be the ratio of the intensity value measured at 1.2 solar-disk radii and the intensity value measured at 
the disk center. 
Figure \ref{fig_simulaFAC} shows the maximum contrast value measured inside the analyzed facular region  for both restored and original images
against the level of stray-light on images. 
The plot illustrates that image restoration enables the facular contrast, at the level of a few percent, to be recovered, and the measurement 
scatter introduced by aberrations to be reduced by approximately one tenth.
 As for sunspots, a slight over-restoration (less than 1$\%$) is observed for the lowest levels of stray-light degradation investigated in this study.

Figure \ref{fig_maxcontr_both} shows the dependence on position inside the solar-disk of the maximum contrast measured for faculae identified 
in the {\it PSPT} and {\it MW} data sets. Results are presented in a similar way to Fig. \ref{fig_spotminint_both}. Besides the already-mentioned 
increase of contrast, we note that the small trend with position inside the solar-disk is not removed by  image 
restoration in 
the analyzed data sets, but appears to 
be slightly enhanced. 
Moreover, the standard deviations inside each bin are increased by image restoration, on average, by a factor 
of 2, which  indicates a wider 
spread in the results. This appears to be in contrast to the results obtained for large-size sunspots identified in the {\it PSPT} data set, for 
which the spread results was reduced remarkably by restoration, and agrees with our results for small sunspots. We comment on these results in the next section. 

\section{Discussion}
   \begin{figure}
   \centering{
\includegraphics[width=7. cm]{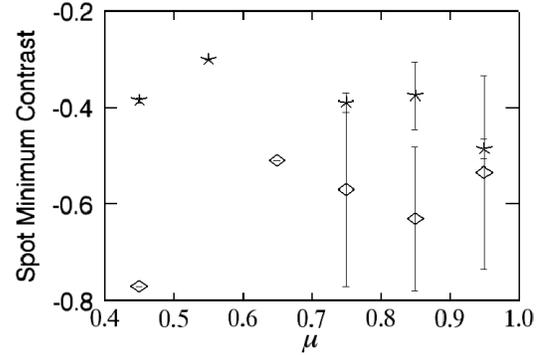}}
       \caption{Minimum contrast of sunspots identified on original (asterisks) and restored (diamonds) {\it test-MW} images.}
         \label{fig_spotminintMWother}
   \end{figure}
%
Our results presented above indicate that restoring improves both spatial resolution and photometric contrast of the images from the {\it PSPT}, 
{\it SD-PSPT}, and {\it MW} data sets.

The increase of the spatial scale on {\it MW} data set is comparable to that derived for {\it PSPT} images   
 (approximately 13$\%$ in both cases). The analyses of photometric properties of solar features  revealed that the increase of the photometry for 
 {\it MW} images is quite poor with respect to that derived for {\it PSPT} data.
The (absolute) increase of contrast of sunspots and facular regions measured in {\it PSPT} images is in fact  60$\%$ and 18$\%$ respectively, while 
the same quantities measured in {\it MW} images are only 6$\%$ and 7$\%$. 
This difference is due to the inefficacy/ineffectiveness of the restoration; this is caused by the lack of information on the intensity 
pattern outside the solar disk image, which is produced by the photographic calibration applied to the {\it MW} images.
As a consequence, measurement of the PSF is affected and the stray-light degradation cannot be compensated for in an accurate way.
To investigate this issue, 
we applied the restoring algorithm to {\it test-MW} images and selected 17 and 19 sunspots on 
original and restored data, respectively. The variation with position on the solar-disk of the 
minimum contrast of these features is shown in Fig.\ref{fig_spotminintMWother}. 
The number of selected sunspots is much smaller than for the {\it PSPT} and {\it MW} images. Sunspots were not selected according to their size; 
the spread of results was larger, and no clear trend with disk position was observed.
 Nevertheless, a comparison with results presented in Fig. \ref{fig_spotminint_both} shows that the compensation for stray-light degradation is 
 larger for {\it test-MW} than  {\it MW} images. In particular, we measured  
a decrease in contrast of approximately $33\%$ for all sunspots, and approximately $40\%$ for sunspots with an umbral area larger than 30 
micro-hemispheres. 
Although smaller, these values are comparable to those measured for the {\it PSPT} data set.
This difference can be partially ascribed to the fact that for high level of stray-light, 
which is true for both {\it test-MW} and {\it PSPT} images,
 the minimization domain (which, in analyzed images, usually extends up to approximately 1.5 solar radii) is not large enough. We  performed 
 some simulations and found that the reduction of the minimization domain produces underestimation of the stray-light 
 and that the effect is larger the larger is the stray-light contamination. 

The analysis of the spread in the measurements allowed the increase of data set homogeneity by restoration to be studied. In the case of the spatial 
resolution, we measured a decrease in the spread of 33$\%$, 14$\%$, and 20$\%$, for the {\it SD-PSPT}, {\it PSPT}, and {\it MW} images, respectively. 
We note that the highest decrease in spread is measured for the data sets with the most homogeneous image content ({\it SD-PSPT}), which suggests that 
the remaining spread measured for the other two data sets is due mainly to variation in the image content discussed in Sect. 4.
In the case of the photometry, measured using  the contrast of solar features, the results are more controversial. We measured a decrease in the 
spread of large-size sunspot contrast for {\it SD-PSPT} and {\it PSPT} data sets. 
On the other hand, 
the spread increased in the contrast between the smallest sunspots and between faculae identified in the {\it PSPT} and {\it MW} data sets.
The increase of spread of facular contrast is due most likely to the enhancement, by restoration, of the dependence of their photometric 
properties to physical characteristics as size, filling factor, presence of bright points and time evolution. 
This is confirmed by the results obtained by analyzing simulated images of the Sun.
Regarding sunspots, \citet{mathew2007} showed that the contrast depends on the size only at the smallest areas. As for faculae, the restoring 
therefore appears to enhance the contrast size-dependence of small sunspots, but reduce the seeing-induced contrast fluctuations of large-size spots.
We are unable to discard the possibility, however, that these enhancements in the contrast spread are generated directly by the restoration method.
In fact,
\citet{walton1999} showed, by the 
analysis of simulated solar images, that the measurement of sunspots contrast is dependent on both their size and position on the solar-disk. 
They showed that the highest quality results are obtained for the largest sunspots (area larger than 100 micro-hemispheres), which  occur at the 
solar-disk center. 
Similarly, \citet{martinez1992}
noticed that potential errors in the assignment of  several parameters used to describe the PSF function play different 
roles at small and large spatial scales. In particular, the degree of stray-light removal on small-scale features
is probably affected by inaccuracies in the parameters of both the Gaussians and the Lorentzian functions of the PSF, while only those concerning 
the Lorentzian part play a role in the quality of the results obtained on 
large-size features. The removal of stray-light effects on contrast 
measurements of small-size features depends on more 
parameters than those acting on contrast measurements of large-size features and is therefore more uncertain. An interesting contribution to this 
topic was provided by \citet{toner1997}.
We note that the uncertainty in the stray-light removal of small-scale features  can  
explain some of the dispersion in the contrast-area relation reported by \citet{mathew2007} for the smallest sunspots that they analyzed.

\section{Conclusions}
We applied an algorithm based on the technique presented by \citet{walton1999} to restore full-disk CaII K solar images for the 
atmospheric and instrumental degradations 
commonly referred to as stray-light. The algorithm was applied to both  current and archive observations, with the aim of 
testing its capability to improve the quality of single images and to recover homogeneous contents of the analyzed data sets. In particular, we 
analyzed images acquired at the Rome Observatory, and two samples of archive  Mt Wilson 
spectroheliograms calibrated by different techniques.  
We found that the application of the restoring algorithm improves both spatial resolution of solar observations, as well as
photometric contrast of sunspots and facular
regions in all data that we analyzed. Results presented in  Sect. 4.1 show that 
the improvement of spatial resolution is comparable for both the present-day  and archive data sets that we analyzed. 
On the other hand, the improvement of photometry for archive data is dependent on the method employed to calibrate the original photographic observations. 
We found  that the estimation of the stray-light level is quite poor in images that, due to the photographic-calibration, contain too little 
information about the intensity 
pattern  outside the solar-disk image. The increase of photometric quality in these images by restoration is 
therefore much smaller than that obtained for both present-day and  differently-calibrated archive observations, as is evident when we compare the 
plots shown in Fig. \ref{fig_spotminint_both} and in Fig. \ref{fig_spotminintMWother}.

Our study indicates that the reduction techniques applied to archive observations have to preserve the largest intensity information 
stored in the original data, even outside the solar-disk image. This allows further improvement of the photometric quality and homogeneity of these observations,
by obtaining quantitative results that are comparable to those produced using current observations. 
Finally, this study confirms, in agreement with \citet{walton1999}, that numerical simulations developed to analyze the recovery of contrast by restoration as a function of feature size and for different PSFs, are required to provide reliable interpretation of photometric
measurements derived from restored full-disk images.

\acknowledgements
The authors thank  S. Solanki and R. Ulrich for useful discussions.  J.A. Bonet and D. Del Moro are acknowledged for reading the 
paper and giving their comments, and the anonymous referee for the helpful 
review of the manuscript.  
The digitization of the Mt Wilson Photographic Archive has been supported by the US National Science Foundation grant ATM/ST 0236682.
This work was supported by the CVS project of Regione Lazio.

\appendix
\section{Original and Restored images}

 \begin{figure*}
   \centering{
\includegraphics[width=6.9cm]{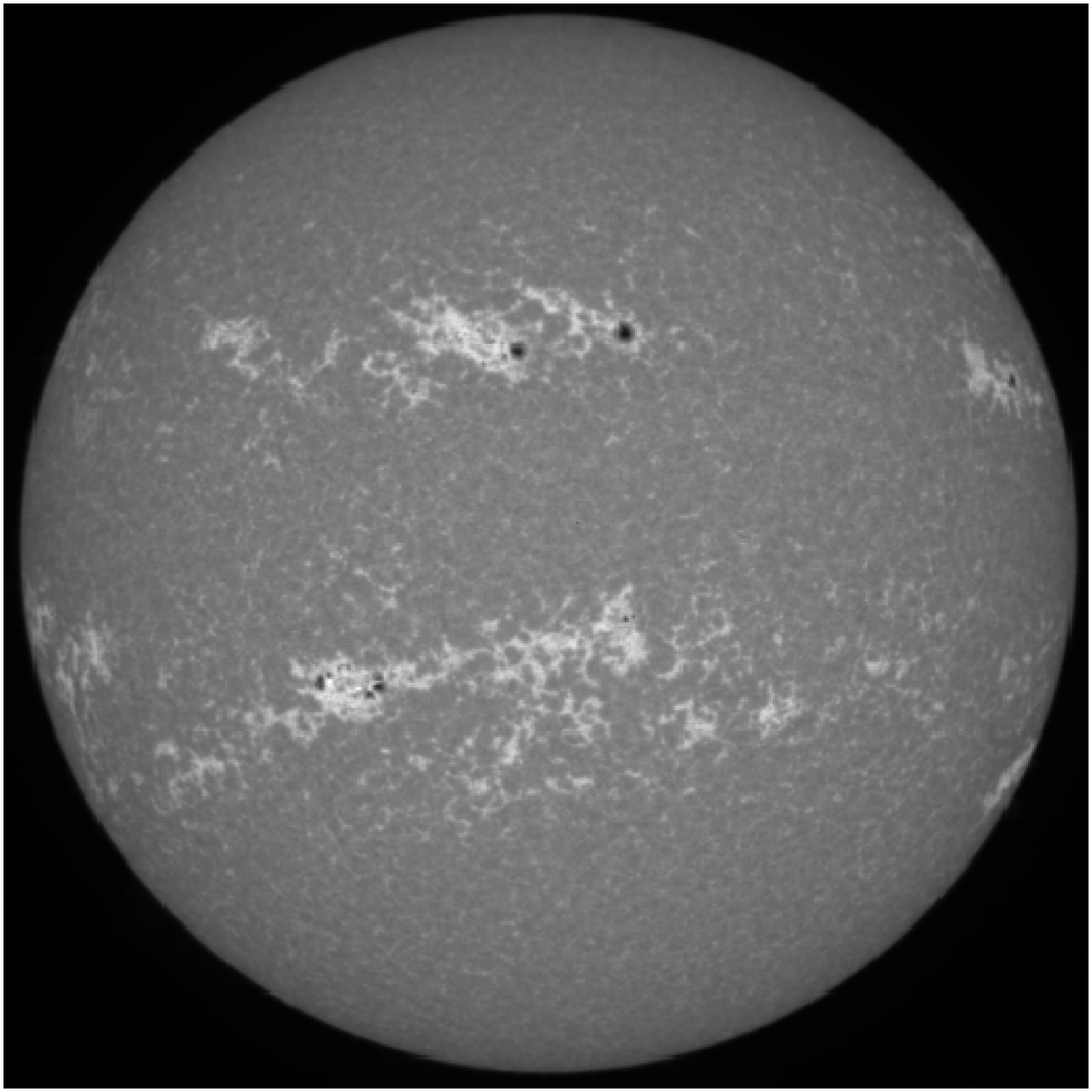}  \includegraphics[width=6.9cm]{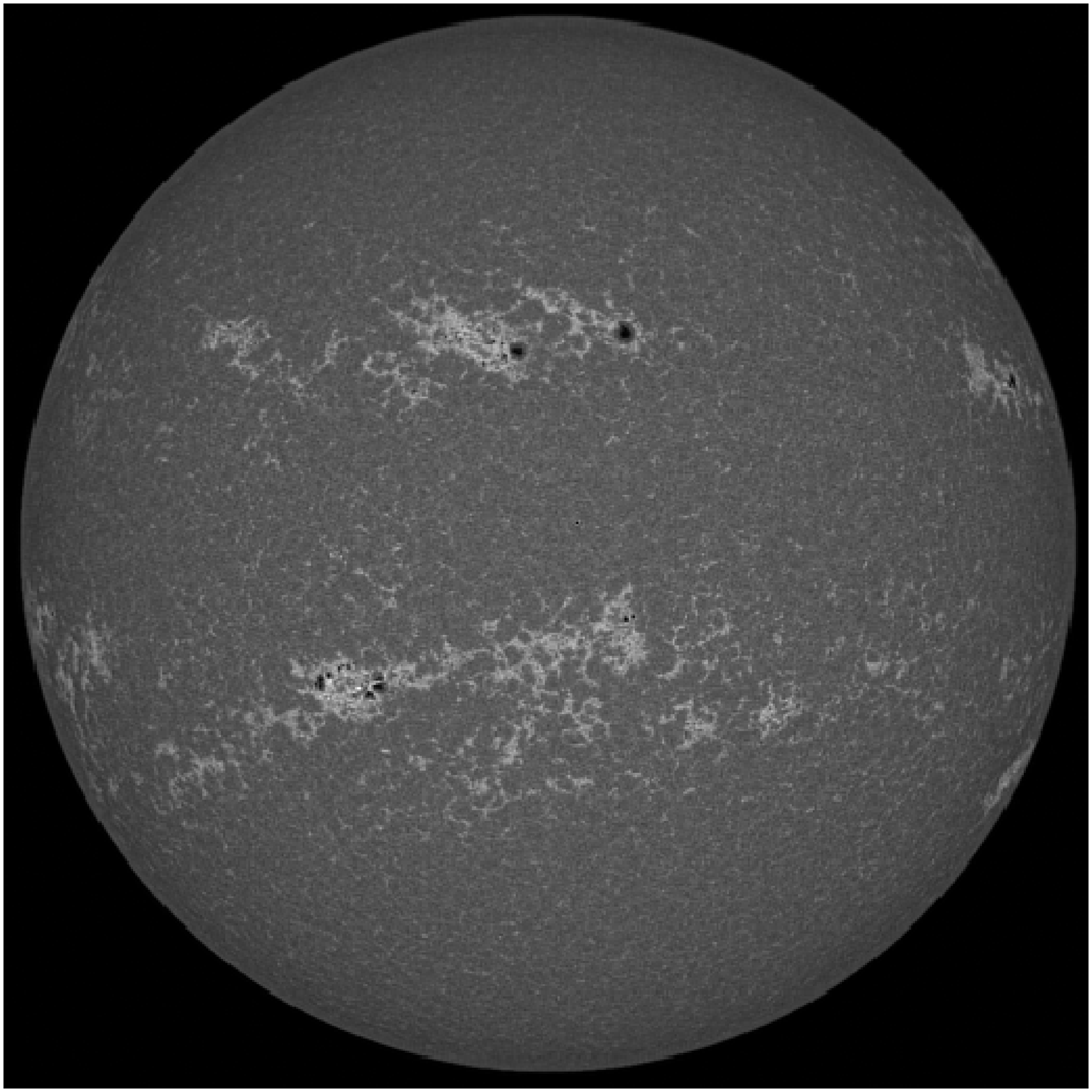}\\
\includegraphics[width=6.9cm]{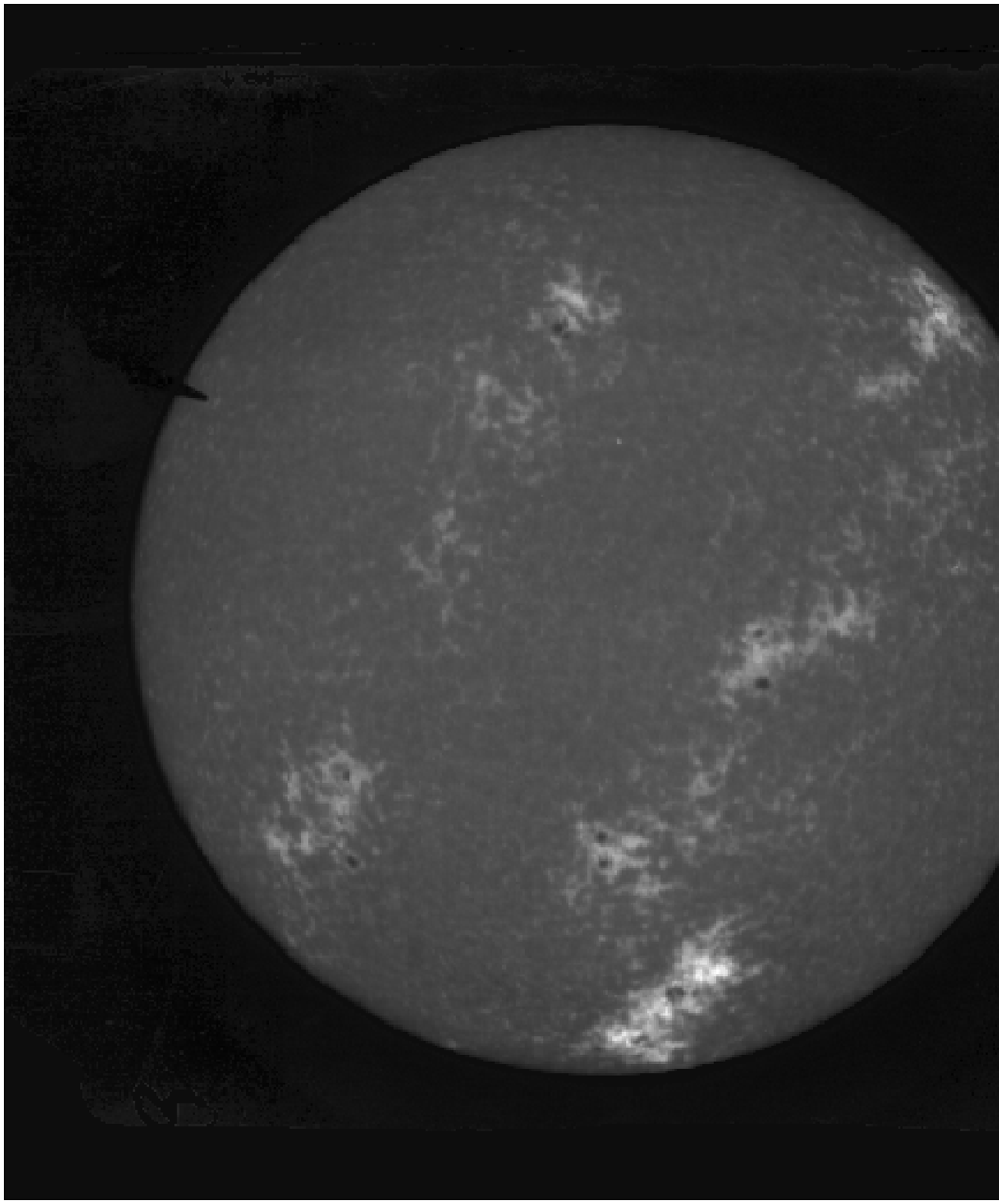} \includegraphics[width=6.9cm]{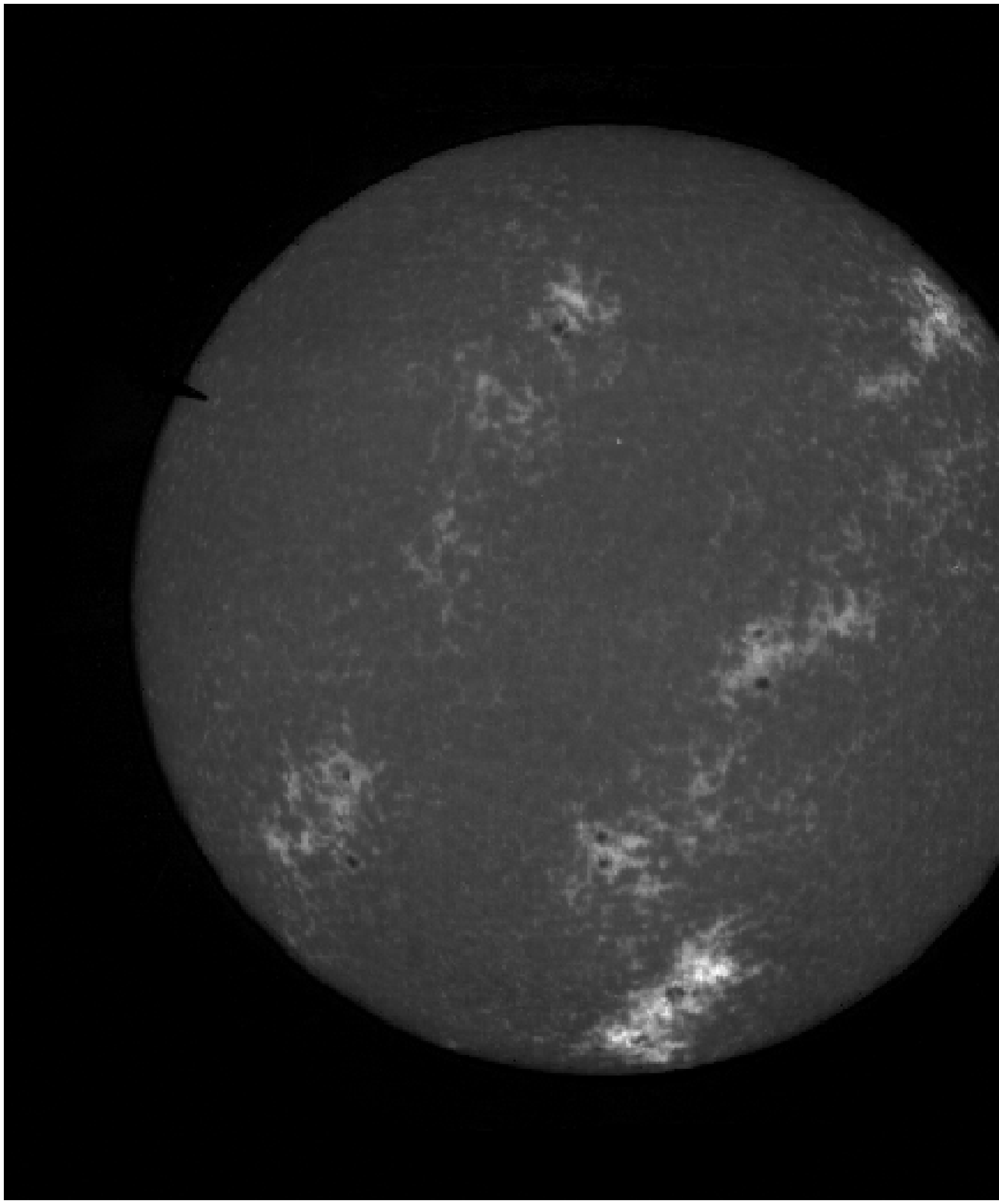}
}
      \caption{Example of original (left) and restored (right) images for the {\it PSPT} (top) and the {\it MW} (bottom) data sets analyzed in this study.
	  The images concern the solar observations taken on July 6, 2000 and on May 29, 1967 at the two sites, respectively.}
        \label{fig_example}
   \end{figure*}

%
   
%
   \begin{figure*}
   \centering{
 \includegraphics[width=7.0cm]{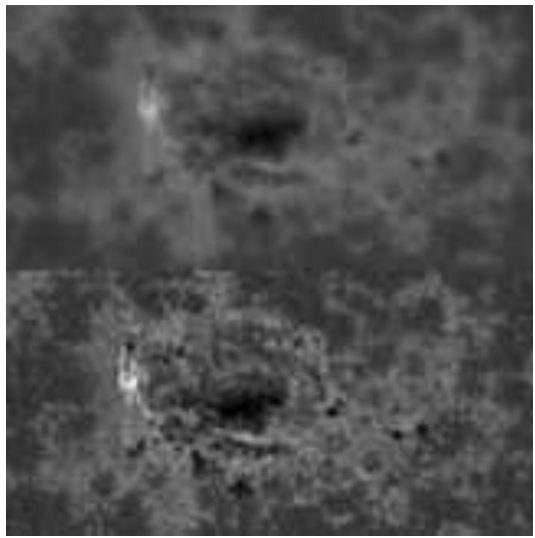}}
      \caption{Details of the analyzed sunspot region AR1038 on an original (top) and restored (bottom) 
{\it SD-PSPT} image. Notice the finer solar features observed in the restored image. Details are given in \S 4.}
         \label{fig_plot_minimt_oneday2}
   \end{figure*}
%

\end{document}